\documentclass[aps,prl,twocolumn,groupedaddress,showpacs,showkeys]{revtex4}
\usepackage{graphicx}
\usepackage{amsmath}
\begin{document}

\title{Small scale lateral superlattices in two-dimensional electron gases prepared by diblock copolymer masks}

\author{S. Hugger}\author{T. Heinzel}\email{thomas.heinzel@uni-duesseldorf.de}
\affiliation{Solid State Physics Laboratory (IPkM), Heinrich-Heine
Universit\"at D\"usseldorf, Universit\"atsstr. 1, 40225
D\"usseldorf, Germany}

\author{T. Thurn-Albrecht }
\affiliation{Institut f\"{u}r Physik, Martin-Luther-Universit\"at Halle-Wittenberg, 06099 Halle, Germany}
\date{\today}

\begin{abstract}
A poly(styrene-block-methylmethacrylate) diblock copolymer in
the hexagonal cylindrical phase has been used as a mask for preparing a periodic gate on top of a $\mathrm{GaAs/Al_xGa_{1-x}As}$-heterostructure.
A superlattice period of $43\,\mathrm{nm}$
could be imposed onto the two-dimensional electron gas. Transport measurements
show a characteristic positive magnetoresistance around zero magnetic
field which we interpret as a signature of electron motion guided by the superlattice potential.
\end{abstract}

\maketitle

\begin{figure}
\includegraphics{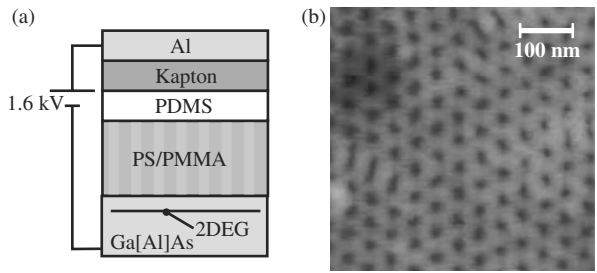}
\caption{(color online) (a) Cross-sectional scheme of the setup used for the orientation of the copolymer film. (b): Topography scan, taken with an atomic force microscope, of the oriented DBC film on top of the Ga[Al]As heterostructure, after removal of the PMMA cylinders. }\label{DBCMaskFig1}
\end{figure}

Lateral superlattices (LSLs) defined in two-dimensional electron gases
(2DEGs) have revealed a wide range of fascinating semiclassical and
quantum transport phenomena, like commensurability resonances, \cite{Ensslin1990,Weiss1991,Lorke1991} Aharonov - Bohm type oscillations
\cite{Nihey1995,Iye2004}, the demonstration of the quasi-particle character of composite fermions, \cite{Kang1993} or the fractal Hofstadter butterfly energy
spectrum. \cite{Schlosser1996,Albrecht2001,Geisler2004} Recent theoretical
suggestions regarding the implementation of spin qubits are based on small-period, high
quality LSLs. \cite{Pedersen2008} For studying LSLs in the quantum regime, a lattice constant comparable to the Fermi wavelength of the electrons, namely in the range of
$30\,\mathrm{nm}$ to $50\,\mathrm{nm}$ in typical
2DEGs, is desirable. However, established lithographic techniques have been
limited to lattice constants above $\approx
80\,\mathrm{nm}$. \cite{Schlosser1996} Patterning techniques based on self-organization have the potential to reduce the LSL period significantly, and diblock copolymers (DBCs) are promising template materials. They offer three potentially useful
phases (spherical, lamellar and hexagonal cylindrical) with lattice constants down to $20\,\mathrm{nm} $ and have already been incorporated in various processing schemes. \cite{Black2007} Moreover, LSLs have already been prepared by transferring the pattern from the spherical phase of a DBC into a
2DEG. \cite{Melinte2004}\\
Here, we report the preparation of a LSL with a period of $\mathrm{a=43\,nm}$ in a 2DEG by using the hexagonal cylindrical phase of
poly(styrene-block-methylmethacrylate) (PS-b-PMMA) to pattern a gate electrode. It is
thereby demonstrated that all the process steps involved are compatible with the established GaAs processing technology.
As compared to the technique of Melinte et al., \cite{Melinte2004} our method is complementary since the copolymer mask has a strictly two-dimensional character with
columns oriented perpendicular to the sample surface and allows
various processing options. Also, in our system the disorder does not seem to dominate over the induced periodic modulation in small magnetic fields. The
pattern transfer into the 2DEG is demonstrated by transport measurements, which show a
characteristic positive magnetoresistance around $B=0$. We support this interpretation by numerical simulations within the Kubo formalism.\\
The experiments have been performed on a commercially available
heterostructure \cite{IntelliEpi} with the heterointerface $65\,\mathrm{nm}$
below the surface. The density and the mobility of
the 2DEG at liquid helium temperatures are $2.3\times 10^{15}\,\mathrm{m^{-2}}$
and $28\,\mathrm{m^{2}V^{-1}s^{-1}}$, respectively.
Hall bars of $10\,\mathrm{\mu m}$ width and a separation of
$13\,\mathrm{\mu m}$ between the voltage probes were prepared by optical lithography and wet
chemical etching. A PS-b-PMMA DBC  with a
molecular weight of $67\,\mathrm{kg/mol}$ of the weight ratio $\mathrm{PS:PMMA = 46:21}$ was dissolved in toluene. DBC films of $\approx 500\,\mathrm{nm}$ thickness were prepared on the preprocessed
heterostructure by spin casting. In the melt, the DBC adopts a
cylindrical morphology with a local hexagonal arrangement of PMMA
cylinders embedded in a PS matrix and lattice constant of
$43\pm 2\,\mathrm{nm}$, in which the orientation of the PMMA cylinders is isotropic. Usage of the DBC film as a lithographic mask requires the cylinders to be oriented perpendicular to the substrate. This was obtained by
annealing the films under an applied electric field of $40\,\mathrm{V/\mu m}$ at $185\,\mathrm{^{\circ}C}$ for $16\,\mathrm{h}$ in nitrogen atmosphere. \cite{ThurnAlbrecht2000,Xu2005} For this purpose, an aluminum layer on top of a kapton foil coated with a crosslinked Polydimethylsiloxane (PDMS) layer of $\approx 20\,\mathrm{\mu m}$ thickness on the side towards the DBC film acted as the counter electrode, \cite{Xu2005} see Fig. \ref{DBCMaskFig1}(a). While this process does not alter the thickness of the DBC film, it imposes a roughness of $\approx 20\,\mathrm{nm}$. The oriented films were exposed to an electron beam with an energy of $20\,\mathrm{keV}$ and a dose of  $50\, \mathrm{\mu C/cm^{2}}$.  The effect of the electrons is to chop the PMMA while at the same time crosslinking the PS. Afterwards, the film was developed in acetic acid for two hours, during which the PMMA chains were removed, leaving behind a PS film with holes in a hexagonal lattice geometry, see Fig. \ref{DBCMaskFig1}(b). The PS matrix shows characteristic deviations from a perfect hexagonal lattice. Namely, the orientation of the lattice vectors varies slowly and occasionally,  grain boundaries are observed. Furthermore, about $1\%$ of the holes are merged with a neighbor.

The gate electrode was electrodeposited from a saturated $\mathrm{CuSO_4}$ solution in water with $25\%$ of methanole added. \cite{ThurnAlbrecht2000} The methanole reduces the capillary forces such that the solution penetrates into the cylindrical holes. A droplet of this solution was placed on top of the PS film and a voltage of $+3\,\mathrm{V}$ was applied to the droplet with respect to the grounded 2DEG, which was accessed via a standard Ohmic contact. The precipitation was stopped as soon as the cylindrical holes were filled and a homogeneous Cu electrode covered the PS film. The fact that Cu could be deposited this way indicates that the holes in the PS film extend all the way down to the GaAs surface.

\begin{figure}
\includegraphics{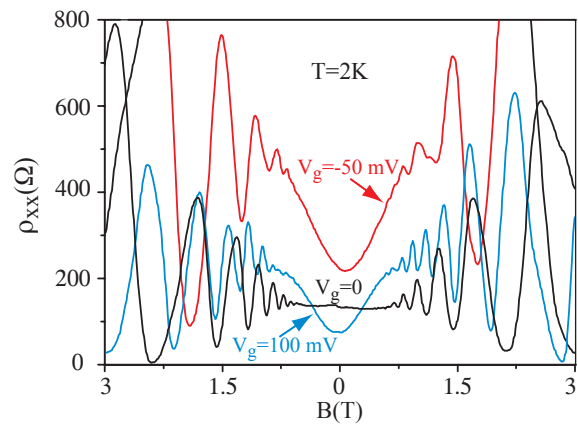}
\caption{(color online) Longitudinal magnetoresistance of the 2DEG underneath the gated region at different gate voltages.}\label{DBCMaskFig2}
\end{figure}

Transport measurements were performed in a $^4$He gas flow cryostat at a temperature of $2\,\mathrm{K}$ using standard lock-in techniques with an
excitation current of $100\,\mathrm{nA}$ at a frequency of
$13.6\,\mathrm{Hz}$.
In Fig. \ref{DBCMaskFig2}, measurements of the longitudinal
magnetoresistance are shown
for different gate voltages. The average electron
density as determined from Hall measurements was  $1.8\times
10^{15}\,\mathrm{m^{-2}}$ at a gate voltage of $V_g= -50\,\mathrm{mV}$ and  $2.9\times
10^{15}\,\mathrm{m^{-2}}$ at $V_g= +100\,\mathrm{mV}$, respectively.

At $V_g\neq 0$, we find a strong, gate voltage dependent
positive magnetoresistance (PMR) around $B=0$ which can extend well above $0.7\,\mathrm{T}$, where Shubnikov-de Haas oscillations set in. Commensurability resonances \cite{Ensslin1990,Weiss1991,Lorke1991} are not observed. The Hall resistivity $\rho_{xy}$ is a straight line and depends only on the average electron density (not shown).
This phenomenology is in tune with measurements on two-dimensional density modulations in square lattices, where the PMR is seen up to $\approx 200\,\mathrm{mT}$ corresponding to the larger LSL periods, while commensurability resonances are observed in various strengths. \cite{Lorke1991,Chowdhury2000,Geisler2005a} From semiclassical simulations, it has emerged that the PMR originates from $\vec{E}\times\vec{B}$ drifts guided along the corrugated potential channels of  the LSL, and breaks down for cyclotron forces larger than the force exerted by the modulation potential, i.e., for $B\gtrsim V_0/(av_F)$ where $V_0$ denotes the modulation potential amplitude and $v_F$ the Fermi velocity. \cite{Grant2000} The strength of the commensurability resonances depends on the amplitude and symmetry of the modulation potential as well as on the disorder scattering. \cite{Grant2000,Mirlin2001,Gerhardts2001}

\begin{figure}
\includegraphics{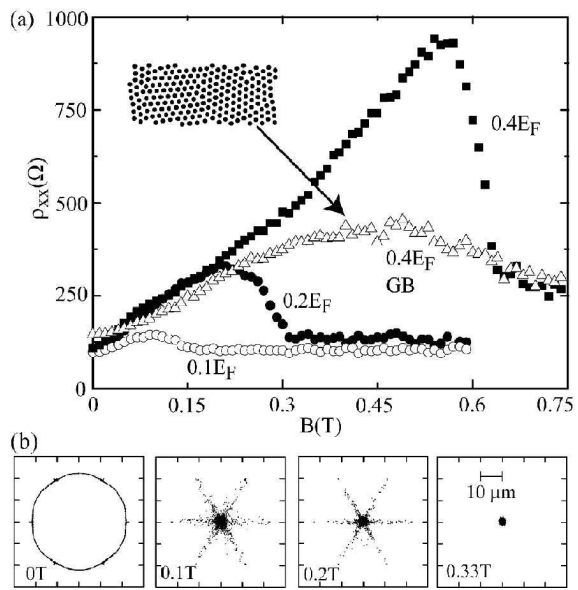}
\caption{(a): Simulated longitudinal resistivity for different modulation
strengths $eV_0$ for a perfect lattice as well as for a LSL with grain boundaries  for $eV_0=0.4E_F$. (b): Final positions of 5000 electrons in different magnetic fields, after $10 \tau_D$ for
$eV_0=0.2E_F$.}\label{DBCMaskFig3}
\end{figure}

To substantiate this interpretation, we performed semiclassical simulations based on
the Kubo formalism. \cite{Kubo1957,Fleischmann1992} The potential $V(x,y)=V_0\cos^2 (k d)$
with $d=\sqrt{(x-x_{c})^2+(y-y_{c})^2}$ and $k=\pi/a$ for
$d<a/2$ and $V=0$ for $d>a/2$ was allocated to every point $(x_c,y_c)$ of the
LSL. For each magnetic field, the semiclassical
equations of motion were solved numerically for 40.000 electrons, and
the conductivity tensor was calculated from
the velocity correlation function via $\sigma_{ij}(B) = \frac{m^* e^2}{\pi {\hbar}^2}
\int\limits_0^\infty \langle v_i(t,B)v_j(0)\rangle e^{-t/\tau_{D}}
\mbox{d}t$, where $t_D=11\,\mathrm{ps}$ and $v_i, i=x,y$ are the Drude scattering time and components of the Fermi velocity, respectively, while $\langle ...\rangle$ denotes emsemble averaging. The longitudinal resistivity $\rho_{xx}(B)$ is then obtained from inversion of the simulated magnetoconductivity tensor.
For a perfect hexagonal LSL with
$a=45\,\mathrm{nm}$, we find a pronounced PMR around $B=0$ which collapses at a magnetic field that increases as $V_0$ increases, see Fig. \ref{DBCMaskFig3}(a).\\
The evolution of the diffusion cloud in increasing magnetic fields provides insight in the origin of this structure, as exemplified for $V_0=0.2E_F$ in Fig. \ref{DBCMaskFig3}(b), where the end positions of 5000 electrons are plotted. They are
injected with velocity $v_F$ from the area $[0,a]\times[0,\sqrt{3}a]$ in random directions after
$110\,\mathrm{ps}$ with disorder scattering turned off. At $B=0$, transport is essentially isotropic.
 As $B$ increases, the electrons are more and more guided along the six directions of high symmetry of the LSL, along which open orbits are possible. \cite{Grant2000,Mirlin2001,Gerhardts2001} Around $B=0.2\,\mathrm{T}$, the guiding collapses and $\rho_{xx}$ experiences a dramatic drop. Commensurability resonances are absent in the simulations as well. \cite{Grant2000}\\
While these simulations nicely reproduce the measured PMR, the sharp drop of $\rho_{xx}$  is not seen experimentally. We attribute this deviation to the disorder in our LSL, as corroborated by magnetotransport simulations in disordered arrays with $V_0=0.4E_F$. A hexagonal lattice containing grain
boundaries was generated by a Monte Carlo method described elsewhere. \cite{Klinkhammer2008} The corresponding simulated trace of $\rho_{xx}(B)$, see Fig. \ref{DBCMaskFig3}(a), shows a significant damping of the magnetoresistance peak, while the slope of the PMR close to $B=0$ remains basically unaffected.\\
In summary, we have generated a  hexagonal LSL with a period of $43\,\mathrm{nm}$ in a 2DEG using diblock copolymer masks in the hexagonal cylindrical phase. A positive magnetoresistance around $B=0$ is found which extends into the Landau quantization regime. It has its origin in electron trajectories which are guided along the corrugated potential channels generated by the LSL. The disorder present in our DBC mask smears the magnetoresistance peak to be expected in perfect  superlattices, but does not destroy the PMR. Since the Fermi wavelength is comparable to the superlattice period in our system, the validity of the semiclassical model is not clear, and it remains to be seen in further work how a quantum model compares to our measurements.\\
Financial support by the Heinrich-Heine-Universit\"at D\"usseldorf and the DFG (SFB 418) is gratefully acknowledged.


\end{document}